\documentclass[aps,twocolumn,showpacs,floatfix]{revtex4}

\usepackage{graphicx}
\usepackage{amsmath}
\usepackage{amssymb}
\usepackage{epsf,latexsym}
\usepackage{times}
\usepackage{bm}
\usepackage[dvips]{color}
\usepackage{float}

\begin{document}
\title{Effect of perturbations on information transfer in spin chains}
\author{R. Ronke$^{1}$
}
\email{rr538@york.ac.uk}
\author{T. P. Spiller$^{2}$
}
\email{t.p.spiller@leeds.ac.uk}
\author{I. D'Amico$^{1}$
}
\email{irene.damico@york.ac.uk}
\affiliation{
$^1$ Department of Physics, University of York, York YO10 5DD, United Kingdom.\\
$^2$ School of Physics and Astronomy, E C Stoner Building, University of Leeds, Leeds, LS2 9JT.\\}
\date{\today}

\begin{abstract}
Spin chains have been proposed as a reliable and convenient way of transferring information and entanglement in a quantum computational context. Nonetheless, it has to be expected that any physical implementation of these systems will be subject to several perturbative factors which could potentially diminish the transfer quality. In this paper, we investigate a number of possible fabrication defects in the spin chains themselves as well as the effect of non-synchronous or imperfect input operations, with a focus on the case of multiple excitation/qubit transfer. We consider both entangled and unentangled states, and in particular the transfer of an entangled pair of adjacent spins at one end of a chain under the mirroring rule and also the creation of entanglement resulting from injection at both end spins.
\end{abstract}


\maketitle

\section{Introduction}

A crucial ingredient in quantum information processing based on solid state systems is the transfer of quantum information. Assuming that there are quantum registers for computing and storing information, the ability to transfer this information reliably and efficiently from one register to another is vital for the construction of larger, distributed and networked systems. A solution to this challenge has been proposed through the use of spin chains \cite{bose2007, kay2009}. The mathematical framework underpinning spin chains can be applied to various physical devices; these could be made of any components whose states can be mapped onto spin $\frac{1}{2}$ particles interacting with their neighbors. Electrons or excitons trapped in nanostructures form explicit examples \cite{damico2007,damico2006,niko2004}, as do nanometer scale magnetic particles \cite{tejada2001} or a string of fullerenes \cite{twamley2003}. Another representation is the encoding into a soliton-like packet of excitations \cite{osborne2004}.

Within spin chains, a single-site excitation is defined as an ``up" spin in a system that is otherwise prepared to have all spins ``down". A discussion about unmodulated spin chains has been given in \cite{bose2003, burgarth2009_2} whereas in \cite{chiara2005} the couplings were chosen to be unequal. There has also been research on wire-like chains with controlled coupling strength at either end \cite{wojcik2005} and transfer through parallel spin chains \cite{burgarth2005}, to name but a few closely related areas. Here we only consider linear spin chains whose coupling strength $J_{i,i+1}$ between two neighboring sites $i$ and $i+1$ has been pre-engineered to ensure perfect state transfer (PST) along the chain \cite{niko2004,chris2005}. For a chain of length $N$ with characteristic coupling constant $J_{0}$, the PST coupling strength sequence is defined as \cite{chris2005}
\begin{equation}
	J_{i,i+1}=J_{0}\sqrt{i(N-i)}.\label{PST}
\end{equation}
For devices based on excitons in self-assembled quantum dots, $J_{0}$ is mainly governed by F\"{o}rster coupling \cite{damico2007}, which in turn depends on the distance between the dots as well as the overlap between the electron and hole wavefunctions in each dot. In gate-defined quantum dots, however, $J_{0}$ will depend on tunnelling and thus on parameters such as the width and height of the barriers which separate the different dots, as well as on the overlap of electronic wavefunctions centered in different dots. For chains of fullerenes or actual atoms $J_{0}$ will represent some ``hopping" parameter describing the propensity of the excitation to transfer from one site to the other. The natural dynamics of a spin chain can then be described by a time independent Hamiltonian as follows
\begin{eqnarray}
\label{hami}
\nonumber{\cal{H}} = \sum_{i=1}^{N}\epsilon_{i}|1\rangle \langle 1|_{i} + \sum_{i=1}^{N-1} J_{i,i+1}[ |1\rangle \langle 0|_{i} \otimes |0\rangle \langle 1|_{i+1} +\\
 |0\rangle \langle 1|_{i} \otimes |1\rangle \langle 0|_{i+1}].
\end{eqnarray}
 In a perfect system (to which perturbations will then be applied) we will assume the single excitation energies $\epsilon_{i}$ to be independent of the site $i$, and therefore only concentrate on the second term of Eq. (\ref{hami}). In some physical systems such as quantum dot strings, $\epsilon_{i}$ could naturally differ according to position, but may be tuned to be the same at all sites via application of local fields \cite{damico2007}. The fidelity $F$, corresponding to mapping the initial state $|\psi_{in}\rangle$ over a time $t$ into the desired state $|\psi_{fin}\rangle$ by means of the chain natural dynamics, is given by
\begin{equation}
	F=|\langle \psi_{fin} |e^{-i{\cal{H}}t/\hbar}| \psi_{ini} \rangle|^{2}, 
\end{equation}
and PST is realized when the evolution is arranged to achieve $F=1$. We use the fidelity of state vectors to determine the transfer quality of information for unentangled states, as detailed for example in \cite{damico2007}. For entangled states, we measure instead the entanglement of formation (EoF) as defined in Ref.~\cite{wootters1998}. \\The time evolution of a system is dependent on its characteristic coupling constant $J_{0}$. In particular, the time scale for PST from one end of a chain to the other, also known as \textit{mirroring time}, is $t_{M}=\pi \hbar/2J_{0}$ so that the periodicity of the system evolution is given by $t_{S}=\pi \hbar/J_{0}$. As the Hamiltonian (\ref{hami}) preserves the excitation number, the evolution of the initial state will remain within the original excitation subspace.

\section{Effect of fabrication defects}

We will now consider the influence of general fabrication defects on linear spin chains with multiple excitations.\\
\textbf{(a) Random noise}\\
We model the effect of fabrication errors (random, but fixed in time) for the energies and couplings in the system by adding to all non-zero entries in the Hamiltonian matrix a random energy $\eta d_{l,m}J_{0}$ for $1 \le l$,$m \le number\,of\,basis\,states$. The scale is fixed by $\eta$ which we set to 0.1 and for each $l \le m$ the different random number $d_{l,m}$ is generated with a flat distribution between zero and unity. For the other side of the diagonal with $m < l$, $d_{l,m}=d_{m,l}$,  preserving the hermiticity of the Hamiltonian. This method of including fabrication defects means that we could observe effects of a reasonable magnitude although clearly other distributions could also be modeled; for specific tests, the weight of the noise would have to be determined according to the individual experiment being simulated.\\
\textbf{(b) Site-dependent ``single-particle'' energies}\\
As a further possible fabrication defect, we consider the effect of the first term of (\ref{hami}) that we previously dismissed under ideal conditions

\begin{equation}
	H_{1} = \sum_{i=1}^{N} \epsilon_{i} |1\rangle \langle1|_{i}.
 \label{enerc}
\end{equation}

$H_{1}$ may represent external perturbations, such as local magnetic fields, or additional single site fabrication imperfections. We thus assume here that $\epsilon_{i}$ is not independent of the site \textit{i} any more.\\
\textbf{(c) Excitation-excitation interactions}\\
In spin chains with multiple excitations, we also consider a perturbation term

\begin{equation}
 \label{interc}
	H_{2} = \sum_{i=1}^{N-1} \gamma J_{0} |1\rangle \langle1|_{i} \otimes |1\rangle \langle1|_{i+1},
\end{equation} 

which represents the interaction between excitations in nearby sites. For example, this may correspond to a biexcitonic interaction in quantum dot-based chains \cite{damico2001, rinaldis2002}.\\
\textbf{(d) Next-nearest neighbor interactions}\\
Finally, we also investigate the effect of unwanted longer range interactions, which could be an issue when considering pseudospins based on charge degrees of freedom. For this we add to (\ref{hami}) the perturbative term
\begin{eqnarray}
\label{hami2}
\nonumber H_{3}= \sum_{i=1}^{N-2} J_{i,i+2}[ |1\rangle \langle 0|_{i} \otimes |0\rangle \langle 1|_{i+2} +\\
 |0\rangle \langle 1|_{i} \otimes |1\rangle \langle 0|_{i+2}].
\end{eqnarray}
The expression for $J_{i,i+2}$ will depend on the type of interaction between spin chain sites.
Here we explicitly consider three cases. 
The first and more general approximates the next-nearest neighbor interaction as proportional to the average of the related interactions between nearest neighbor sites, 
\begin{equation}
 \label{Ji+2}
	J_{i,i+2}=\Delta(J_{i,i+1}+J_{i+1,i+2})/2,
\end{equation} 
with the parameter $\Delta$ defining the strength of the interaction.
This expression simulates the original coupling modulation of the chain. 

Secondly we explicitly consider dipole-dipole interactions, which are relevant, e.g. to chains of quantum dots with exciton qubits and  F\"{o}rster coupling~\cite{spiller}. In this case the coupling between sites scales as $1/R^3$, with $R$ the distance between the two sites considered. For roughly equidistant sites, we then expect the next-nearest neighbor couplings to be about a tenth of the nearest neighbor couplings.
By using this and  Eq.~(\ref{PST}) we obtain
\begin{equation}
 \label{Ji+2dip-dip}
	J^{dip}_{i,i+2}=J_0\left\{[i(N-i)]^{-\frac{1}{6}}+[(i+1)(N-i-1)]^{-\frac{1}{6}}\right\}^{-3}.
\end{equation}
Finally we consider the case of coupling due to tunnelling, relevant e.g. to graphene quantum dots with spin qubits~\cite{trauzettel2007}. Here the coupling scales as  $4t^{2}/U$, with $U$ the on-site Coulomb energy and $t\propto R e^{-R|k|}$ the tunnelling (hopping) parameter, with $k$ the 'forbidden' momentum in the barrier~\cite{trauzettel2007}. In this case there is no explicit expression for the next-nearest neighbor couplings in terms of $\{J_{i,i+1}\}$, but $\{J^{tun}_{i,i+2}\}$ can be determined numerically by using the expression for $t$ and Eq.~(\ref{PST}). We note that we expect $\{J^{tun}_{i,i+2}\}$ to be very small, as the interaction decays exponentially with the distance.

\begin{figure}
 \centering
 \includegraphics[width=0.45\textwidth]{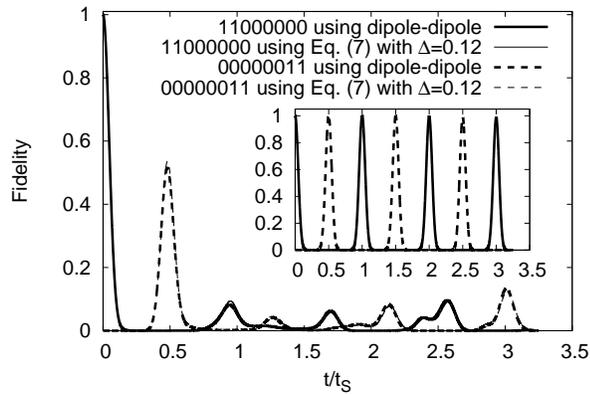}
 \caption{Influence of $H_{3}$ on the example of an 8-spin chain with initial input state $|\psi_{in}\rangle = |11000000\rangle$, fidelity vs. rescaled time $t/t_{S}$, according to both Eq. (\ref{Ji+2}) and Eq.~(\ref{Ji+2dip-dip}) (dipole-dipole coupling). The peak at $t=0.5t_{S}$ is approximately 0.52. As the model of Eq. (\ref{Ji+2}) matches the data derived from Eq. (\ref{Ji+2dip-dip}) extremely well, the lines of the respective plots are nearly indistinguishable.\\Inset: As for main panel but for graphene quantum dots and spin qubits, with tunnelling coupling (thin lines) and coupling according to Eq. (\ref{Ji+2}), $\Delta = 0.0001$ (bold lines). Again results from the two models for the coupling constants are almost indistinguishable from each other.}
 \label{fig:NNNIcomp}
\end{figure}

If we consider chains of quantum dots with exciton qubits, we can assume $J_{max}\approx 1 meV$,  where $J_{max}=max_{i}\{J_{i,i+1}\}$, and $R_{min}\approx 10 nm$. Results in Fig. \ref{fig:NNNIcomp} then show that this is in very good agreement with considering  Eq. (\ref{Ji+2}) with $\Delta = 0.12$, a value that one would expect from dipole-dipole interaction and basically equidistant sites (see discussion above). In this case the effect of next-nearest neighbor interaction is extremely detrimental to the system as even the fidelity of the first state transfer at $t=0.5 t_{S}$ is reduced by almost 50\%.

By contrast, the tunnelling mechanism case leads to very different results. Using the parameters in \cite{trauzettel2007} for graphene dots and spin qubits, we obtain the results shown in the inset of Fig. \ref{fig:NNNIcomp}: after $3t_{S}$ the system has lost less than 2\% of its fidelity. These results are in very good agreement with Eq. (\ref{Ji+2}) with $\Delta=0.0001$ as can be seen in the inset of Fig.1, where the plots are virtually indistinguishable. It is very encouraging that in this case realistic parameters point to such small values of $\Delta$ and thus generate very high fidelity transfer.

In the following we will use the expression in Eq. (\ref{Ji+2}) to further discuss the effects of $H_{3}$. As values of $\Delta$ smaller than 0.01 have a minor detrimental effect on the system, we will from here onwards focus on the range $0.01 \leq \Delta \leq 0.1$, where the upper bound may be of interest to some experimental implementations, e.g. as discussed in the dipole-dipole interaction case.

We will now consider the effect of the fabrication defects (a) to (d) first on the transfer of factorisable states, i.e. unentangled chains, and then on entanglement creation and entanglement transfer along a spin chain. For unentangled states, we will consider a 6-spin chain for all investigations of fabrication defects while in the case of entanglement, we also consider an 8-spin chain. As we will explicitly see later for $H_{1}$, the influence of fabrication defects does depend on the chain length, but not on the parity of the chain.

\subsection{Transfer of unentangled states}

The device we consider in this paper is a linear spin chain with couplings fixed such that the conditions for PST are satisfied. One of the properties of these devices, which we make heavy use of, is the mirroring rule \cite{albanese2004, karbach2005}. The mirroring rule is such that any state injected into a linear spin chain, subject to the constraints of Eq. (\ref{PST}) and site-independent $\epsilon_{i}$, evolves into its ``mirror state'', where the symmetry center of a chain is the middle point in chains with $N$ even, or middle site $({N+1})/{2}$ in chains with $N$ odd. This mirroring property is independent of the number of excitations a state comprises and also of the length of the chain.  Furthermore the mirroring rule holds for states spread across the whole chain: all excited sites are mirrored across to their ``twin'' with respect to the chain center of symmetry. We call this mirrored state the ``twin state''. When investigating the quality of a device with fabrication defects of any sort, we are primarily interested in the effect on the ``twin state'' at $t=t_{M}$ but it is also important to consider the effects on the next few periods as some quantum information protocols may require periodicity of their systems \cite{clark2005}.

\begin{figure}
 \centering
 \includegraphics[width=0.45\textwidth]{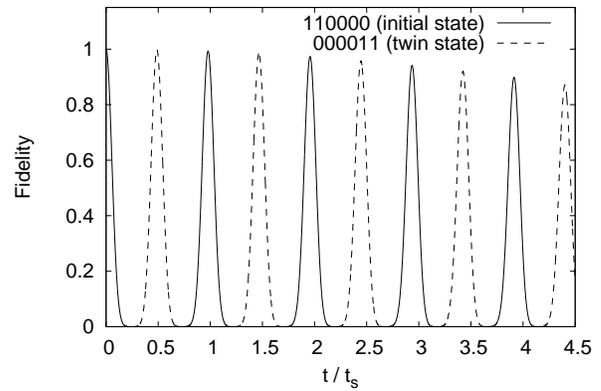}
 \caption{Influence of random noise (see (a) in text) for $\eta=0.1$ on a 6-spin chain with two excitations, fidelity vs. rescaled time $t/t_{S}$. States other than the input state and its twin state are not shown. The first peak at $t=0.5t_{S}=t_{M}$ is 0.9975.}
 \label{fig:6spinnoise}
\end{figure}

 As can be seen in Fig. \ref{fig:6spinnoise}, the effect of random noise leads to a continuous and definite decline in the transfer fidelity of both the input state $|110000\rangle$ and its ``twin state`` $|000011\rangle$ at the mirroring times. Naturally, this means an increased probability of the occurrence of other possible states (not shown on the graph) but this happens on a relatively unpredictable basis, with no one state ever becoming and remaining particularly prominent.

\begin{figure}
 \centering
 \includegraphics[width=0.45\textwidth]{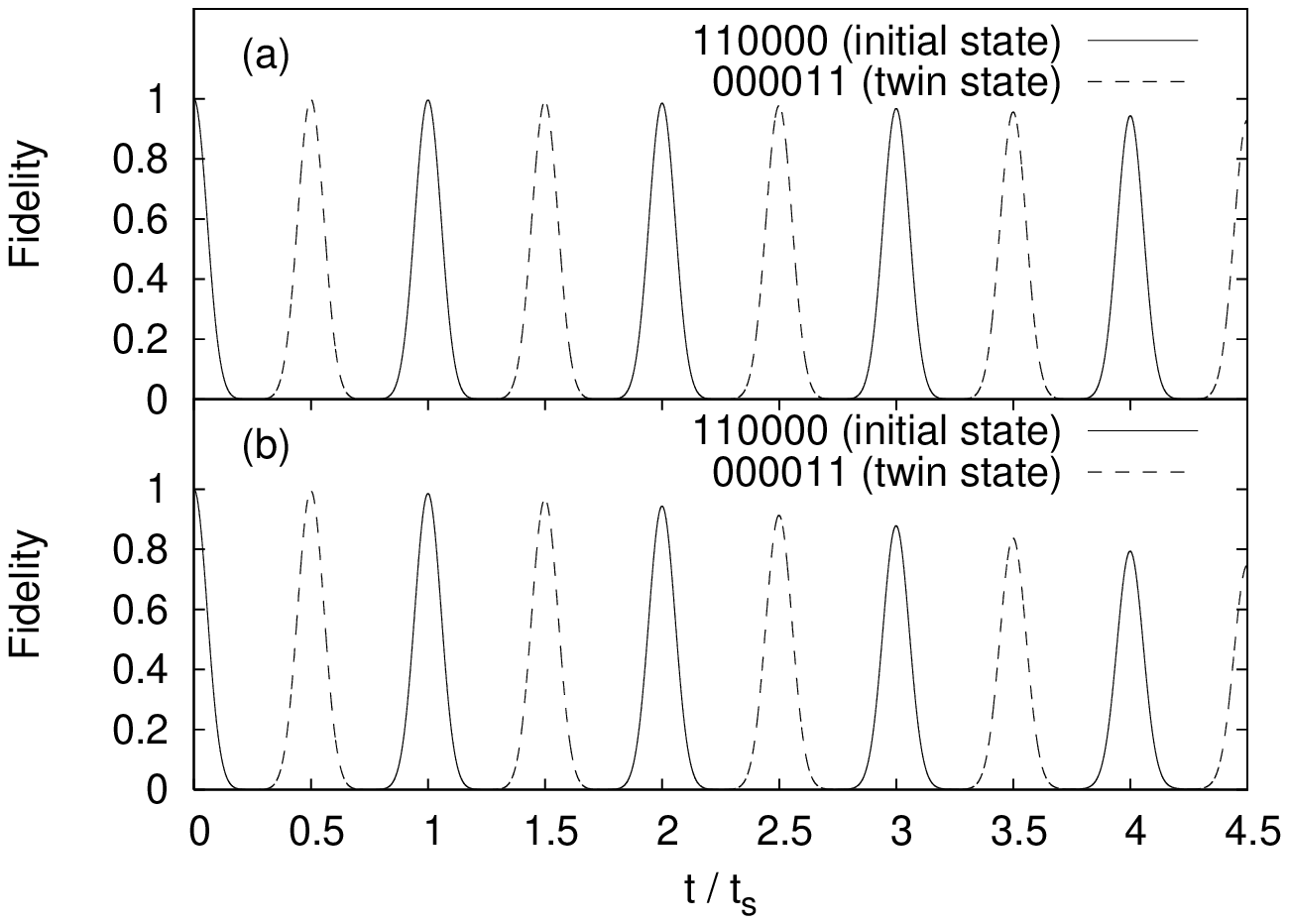}
 \caption{Effect of perturbation $H_{2}$ (Eq. (\ref{interc})) on a 6-spin chain, fidelity vs. rescaled time $t/t_{S}$.\\(a) $\gamma=0.05$: the first peak at $t=0.5t_{S}=t_{M}$ is 0.9988.\\(b) $\gamma=0.1$: the first peak at $t=0.5t_{S}=t_{M}$ is 0.9954.}
 \label{fig:interc}
\end{figure}

\begin{figure}
 \centering
 \includegraphics[width=0.45\textwidth]{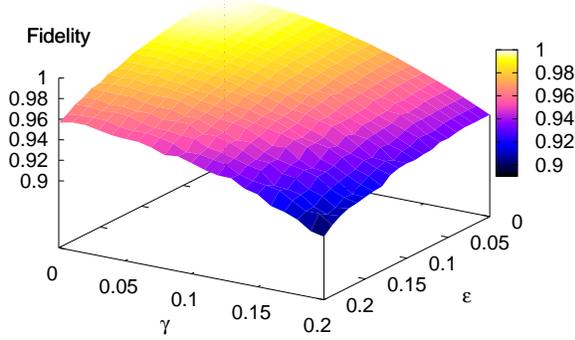}
 \caption{(Color online) Revival fidelity of the state $|\psi_{in}\rangle = |111000\rangle$ in a linear 6-spin chain, measured on the first revival peak at time $t=t_{S}$, vs $\gamma$ and $\epsilon$ .}
 \label{fig:3d}
\end{figure}

In comparison, when looking at the evolution over a few periods, even a relatively large value of $\gamma=0.05$ in Eq. (\ref{interc}) has less effect on the 6-spin chain, as shown in Fig. \ref{fig:interc} (a). Similarly, the on-site energies represented by Eq. (\ref{enerc}) also lead to an unrecoverable decay in the state transmission fidelity, with the excitations being ultimately entirely spread out along the chain (not shown). A combination of any of these three perturbation factors simply accelerates the decay trend of the desired states. \\However, if we simply analyze the first revival peak at $t=t_{S}$, we see that for varying  $\gamma$ and $\epsilon$, even when perturbing the system by as much as 20\% of $J_{0}$, the system suffers less than a 10\% loss in fidelity (Fig. \ref{fig:3d}). Here, to represent the fact that $\epsilon_{i}$ is site-dependent, we used a value $\epsilon_{i} = \epsilon J_{0} d_{i}$ for all sites \textit{i}, with $0 \le d_{i} \le 1$ a random number generated from a flat distribution for each $i$. An average over 200 realizations for every value of $\epsilon$ was then taken.

\begin{figure}
 \centering
 \includegraphics[width=0.45\textwidth]{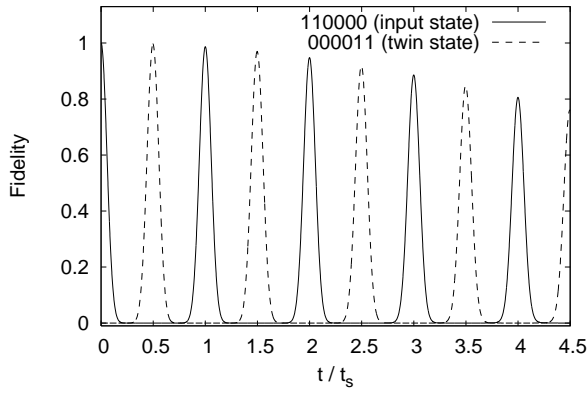}
 \caption{Influence of $H_{3}$ (Eq.(\ref{hami2})) with $\Delta=0.01$: fidelity of the state $|\psi_{in}\rangle = |110000\rangle$ in a 6-spin chain vs. rescaled time $t/t_{S}$. The first peak at $t=0.5t_{S}=t_{M}$ is 0.9966.}
 \label{fig:110000nnni}
\end{figure}

Finally, we also note that next-nearest neighbor interactions perturb the system similarly to noise. This can be seen in Fig. \ref{fig:110000nnni} where even for a reasonable value of $\Delta=0.01$, the transfer peaks of both the initial state of a 6-spin chain and its mirror twin quickly decay. In order to achieve the same fidelity loss at $t=4.5 t_{S}$ as for $\Delta=0.01$ (Fig. \ref{fig:110000nnni}), we have to consider a value of $\gamma$ {\it one order of magnitude bigger}, as is shown in Fig. \ref{fig:interc} (b).

\begin{figure}
 \centering
 \includegraphics[width=0.48\textwidth]{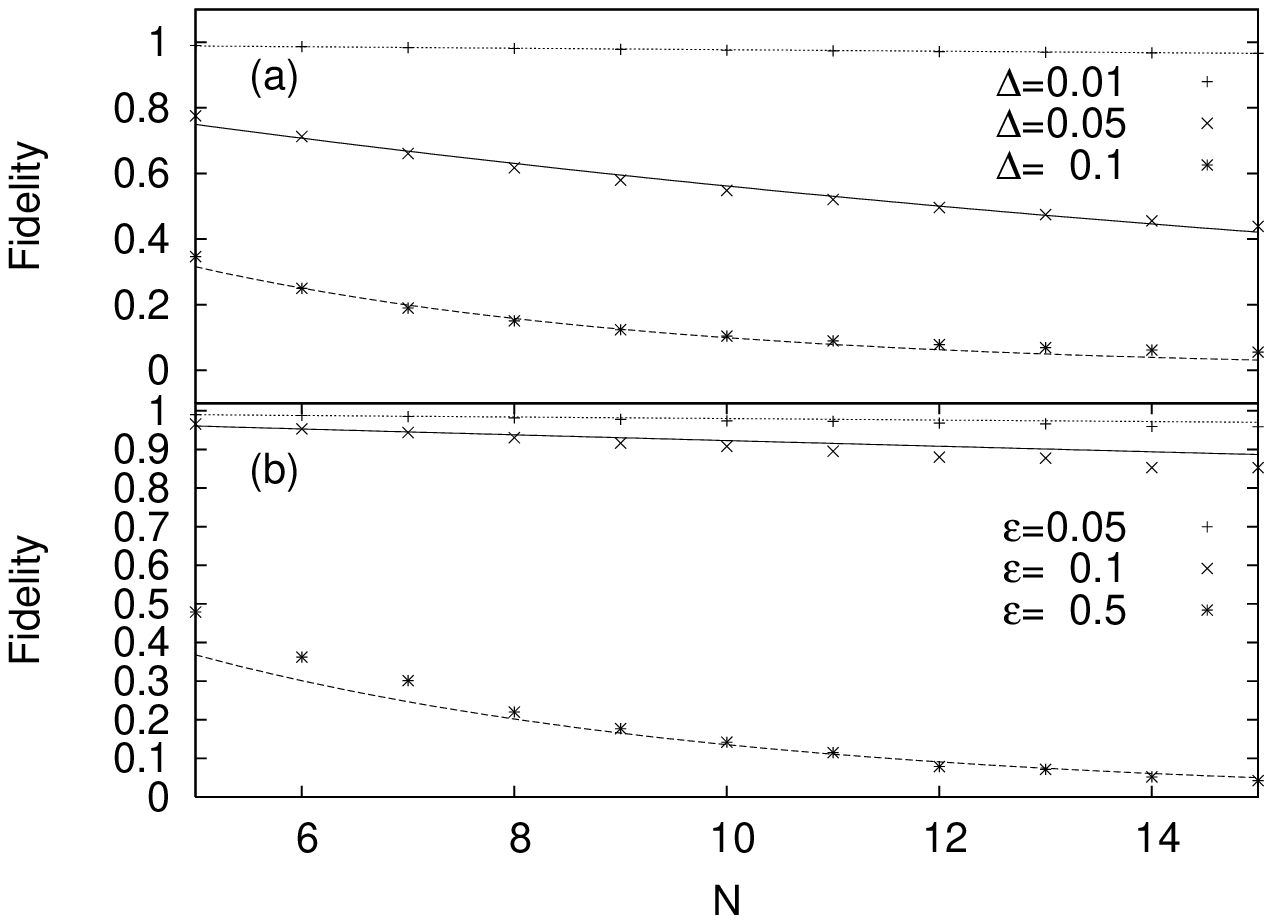}
\caption{(a) Fidelity of initial vector $|110\cdots0\rangle$ at $t=t_{S}$ vs. chain length $N$ for three values of $\Delta$, as labelled. Fits are according to $e^{-N \Delta^{2}/\Delta^{2}_{0}}$ with $\Delta_{0}=0.21$.\\(b) Average fidelity of initial vector $|110\cdots0\rangle$ at $t=t_{S}$ vs. chain length $N$ for three values of $\epsilon$, as labelled. Fits are according to $e^{-N \epsilon^{2}/\epsilon^{2}_{0}}$ with $\epsilon_{0}=1.12$.}
 \label{fig:110000_nnnienerc}
\end{figure}

In our simulations we have kept the value of $J_{max}$  constant as $N$ is varied: this models the physical constraint that in any realistic system the coupling strength is capped by a maximum characteristic value. As a result $J_{0}=2J_{max}/N \: (J_{0}=2J_{max}/N \sqrt{1-1/N^{2}})$ for even (odd) chains. To avoid this implicit dependence on $N$, we have here set $\epsilon_{i} = \epsilon J_{max} d_{i}$ (where $d_{i}$ is a random number from a uniform distribution within 0 and 1) and averaged every point on the graph from 200 random realizations. As we see in part (a) of Fig. \ref{fig:110000_nnnienerc}, the effect of $H_{3}$ becomes very detrimental to the system even for the relatively small value of $\Delta=0.05$, although the loss of fidelity for $\Delta=0.01$ is very small even for long chains. On the other hand, the effect of on-site energies may be tolerable for values of $\epsilon$ up to $0.1$, where long chains of $N=15$ suffer less than a 20\% loss in fidelity.

A simple estimate of the effect of errors in the energy level spectrum \cite{chris2005} suggests an overall error, or loss in fidelity for PST that scales as an exponential decay in $N$ with Gaussian dependence on the characteristic noise parameters. We compare this with numerical results in Fig. \ref{fig:110000_nnnienerc}, where the loss of fidelity due to $H_{1}$ scales as $e^{-N f(\epsilon)}$ and the loss due to $H_{3}$ scales as $e^{-N f(\Delta)}$ with increasing chain length $N$, where $f(\Delta)=\Delta^{2}/\Delta^{2}_{0}$ and $f(\epsilon)=\epsilon^{2}/\epsilon^{2}_{0}$ so that $\Delta_{0}$ and $\epsilon_{0}$ characterize the impact of the noise. The comparison in Fig. \ref{fig:110000_nnnienerc} shows that this simple analytical form can reproduce the numerical results to a high degree of accuracy: the loss of fidelity scales indeed as an exponential decay with Gaussian damping in the noise parameters.

We conclude from these results that the transfer of unentangled states across the device is very robust against the perturbations $H_{1}$ and $H_{2}$, and less so for $H_{3}$. The effect of noise in the system, as we have implemented it, is slightly more noticeable at the transfer time $t_{M}$ but still allows for excellent transfer at a loss rate of just over 10\% over the course of 4 periods. With regards to longer term periodicity, it is the next-nearest neighbor interaction term $H_{3}$ which perturbs the system most for the values of $\epsilon_{i}$, $\gamma$ and $\Delta$ shown.

\subsection{Transfer and creation of entangled states}

One of the most outstanding properties of spin chains is their ability to not just transfer reliably factorisable states, but also to transfer information encoded as entangled states. This is again based on the mirroring rule \cite{albanese2004, karbach2005} and was first mentioned and discussed under various aspects in Refs.~\cite{plenio2004, maru2007, srini2007, cubitt2008, tsomokos2007, chris2004, chris2005}. Entanglement is one of the key resources in quantum computing, and is crucial to some quantum cryptography protocols, and to quantum teleportation. Being able to reliably transfer entanglement from one place to another is therefore a core interest that the device we are analyzing should be able to respond to. This property follows from the fact that the set of states $|\phi_{\{i\},n}\rangle=|i_1,i_2,\dots,i_N\rangle$ , $i_j=0,1$ to which the mirroring rule applies, is a basis set so that any state, and in particular entangled states, can be written as the superposition $|\Psi\rangle=\sum_{\{i\},n} c_{\{i\},n} |\phi_{\{i\},n}\rangle$. Here $n$ is the total number of excitations in the state, and $\{i\}$ the ensemble of indices $i_{j}$ different from zero. The above relation implies that after $t_{M}$ has passed the following ''twin state`` is reached: $|\Psi_{twin}\rangle=\sum_{\{i\},n} c_{\{i\},n} \exp\{-iHt_{M}\}|\phi_{\{i\},n}\rangle = \sum_{\{i\},n} c_{\{i\},n} |\phi_{\{i\},n;twin}\rangle$, so that in particular any entangled state is transferred into its mirror entangled state.

When considering spin chains with (i) an entangled initial state or (ii) an initial state that leads to entanglement, it is more useful to observe the evolution of the EoF in the system. We use chains with an initial Bell state on spins 1 and 2 to represent scenario (i), as for example $|\psi_{in}\rangle = \frac{1}{\sqrt{2}} (|1000\rangle+|0100\rangle)$. Accordingly, we monitor the EoF in the reduced density operator of spins $(1,2)$, tracing out the rest of the chain.  To monitor the ``twin state" entanglement we calculate the EoF for the reduced density operator of spins $(N-1,N)$. Scenario (ii) is different in that the initial state of the chain is not entangled, but will lead to entanglement through natural dynamics. As an example of this, we use a linear chain with input $|+\rangle = \frac{1}{\sqrt{2}} (|0\rangle + |1\rangle)$ on both spins 1 and $N$. This is equivalent to an initial state $|\psi_{in}\rangle = |0_{1}0_{N}+1_{1}0_{N}+0_{1}1_{N}+1_{1}1_{N}\rangle \otimes |0_{2} 0_{3} \cdots 0_{N-1}\rangle$, where the subscripts designate the spin site. A system set up in this way will then show a maximally entangled state in spins 1 and $N$ at time $t=t_{M}$ \cite{clark2005, clark2007, yung2005, yung2006}.

\begin{figure}
 \centering
 \includegraphics[width=0.45\textwidth]{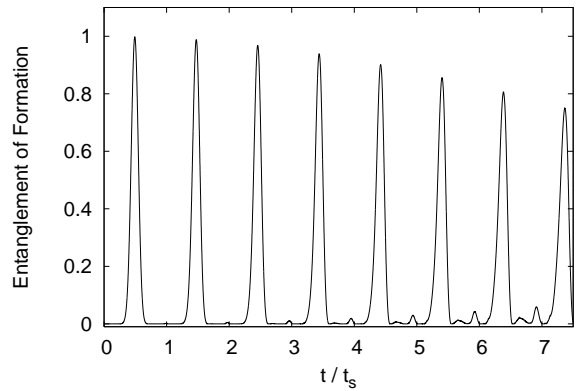}
 \caption{Influence of random noise (see (a) in text) for $\eta=0.1$ on the entanglement of formation between the end spins of a linear 8-spin chain with input state $|\psi_{in}\rangle = 1/2(|0_{1}0_{8}\rangle+|1_{1}0_{8}\rangle+|0_{1}1_{8}\rangle+|1_{1}1_{8}\rangle)\otimes |0_{2} \cdots 0_{7}\rangle$ vs. rescaled time $t/t_{S}$. The first peak at $t=0.5t_{S}=t_{M}$ is 0.9983.}
 \label{fig:gatenoise}
\end{figure}

\begin{figure}
 \centering
 \includegraphics[width=0.45\textwidth]{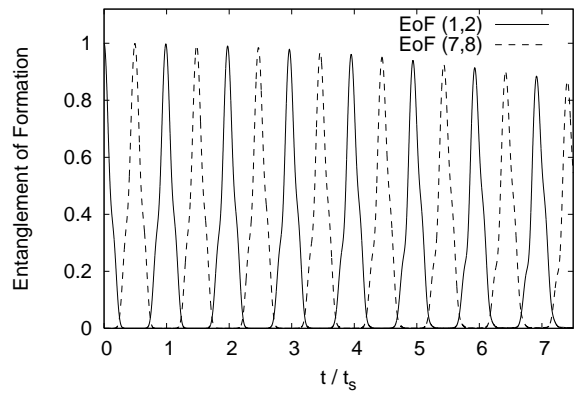}
 \caption{Influence of random noise (see (a) in text) for $\eta=0.1$ on the entanglement of formation between the end spins of an 8-spin chain with input state $|\psi_{in}\rangle = 1/\sqrt{2}(|10000000\rangle + |01000000\rangle)$ vs. rescaled time $t/t_{S}$. The first peak at $t=0.5t_{S}=t_{M}$ is 0.9993.}
 \label{fig:0110noise}
\end{figure}

The effect of noise on entangled chains shows a similar trend to that of unentangled chains. However, we notice that the loss of EoF in Fig. \ref{fig:gatenoise} which represents case (ii) is nearly 10\% bigger than the loss of EoF in Fig. \ref{fig:0110noise} (case (i)) over the course of the shown 7 periods. As a comparison, the unentangled state of Fig. \ref{fig:6spinnoise} loses a similar amount of fidelity over the same amount of time as the chain in Fig. \ref{fig:gatenoise}.

\begin{figure}
 \centering
 \includegraphics[width=0.45\textwidth]{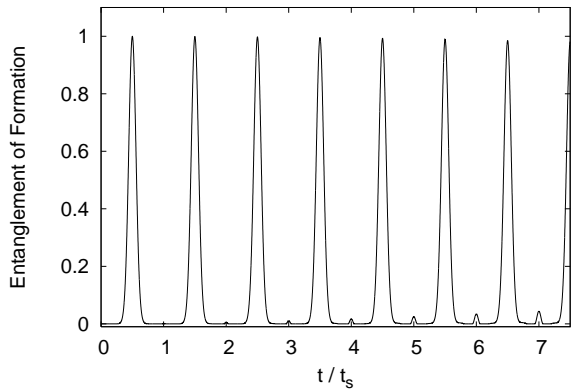}
 \caption{Influence of $H_{2}$ (Eq.(\ref{interc})) with $\gamma=0.05$ on the entanglement of formation between the end spins of a linear 8 spin chain with input state $|\psi_{in}\rangle = 1/2(|0_{1}0_{8}\rangle+|1_{1}0_{8}\rangle+|0_{1}1_{8}\rangle+|1_{1}1_{8}\rangle)\otimes |0_{2} \cdots 0_{7}\rangle$ vs. rescaled time $t/t_{S}$. The first peak at $t=0.5t_{S}=t_{M}$ is 0.9996.}
 \label{fig:gateinterc}
\end{figure}

There is no effect from perturbation $H_{2}$, (Eq. (\ref{interc})), in case (i), as there is only one excitation in the system in both amplitudes. Obviously this would not be so for a Bell state with a doubly excited amplitude. Similarly, the effect of $H_{2}$ in case (ii) is restricted to the two excitation subspace populated by the evolution of the $|1_{1}1_{N}\rangle \otimes |0_{2} \cdots 0_{N-1}\rangle$ component of $|\psi_{in}\rangle$ only and is thus not very prominent, as is shown in Fig. \ref{fig:gateinterc}.

\begin{figure}
 \centering
 \includegraphics[width=0.48\textwidth]{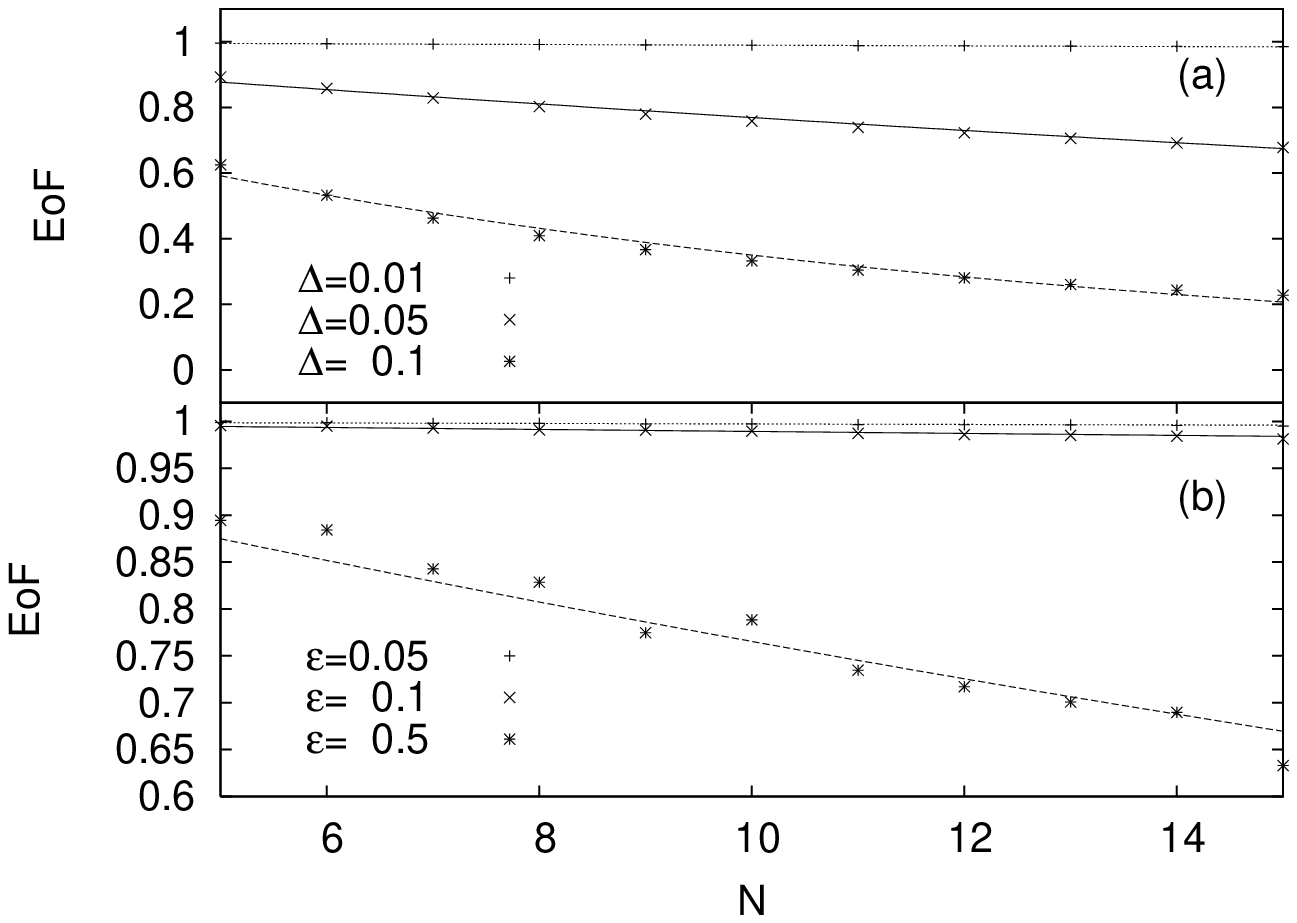}
\caption{For chains with type (i) entanglement (spins 1 and 2 are entangled at $t=0$):\\ (a) EoF of qubits 1 and 2 at $t=t_{S}$ vs. chain length $N$ for three values of $\Delta$, as labelled. Fits are according to $e^{-N \Delta^{2}/\Delta^{2}_{0}}$ with $\Delta_{0}=0.31$.\\(b) Average EoF of qubits 1 and 2 at $t=t_{S}$ vs. chain length $N$ for three values of $\epsilon$, as labelled. Fits are according to $e^{-N \epsilon^{2}/\epsilon^{2}_{0}}$ with $\epsilon_{0}=3.06$.}
 \label{fig:10+01_nnnienerc}
\end{figure}

\begin{figure}
 \centering
 \includegraphics[width=0.48\textwidth]{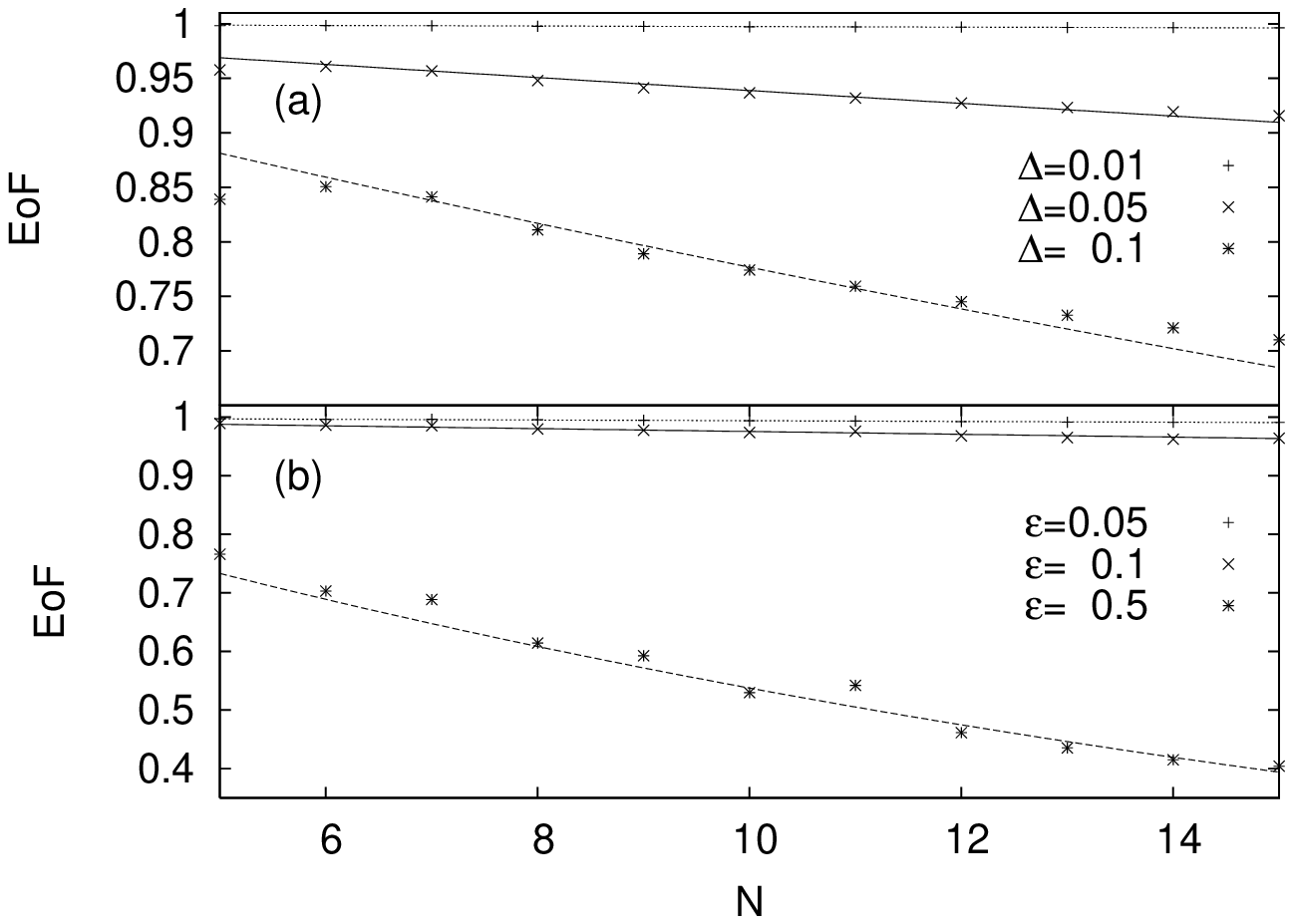}
\caption{For chains with type (ii) entanglement (spins 1 and N become maximally entangled at $t=t_{M}$):\\ (a) EoF of qubits 1 and $N$ at $t=t_{M}$ vs. chain length $N$ for three values of $\Delta$, as labelled. Fits are according to $e^{-N \Delta^{2}/\Delta^{2}_{0}}$ with $\Delta_{0}=0.63$.\\(b) Average EoF of qubits 1 and $N$ at $t=t_{M}$ vs. chain length $N$ for three values of $\epsilon$, as labelled. Fits are according to $e^{-N \epsilon^{2}/\epsilon^{2}_{0}}$ with $\epsilon_{0}=2.01$.}
 \label{fig:gate_nnnienerc}
\end{figure}

The effect of on-site energies on the other hand remains, regardless of the system. We demonstrate this in part (b) of Figs. \ref{fig:10+01_nnnienerc} and \ref{fig:gate_nnnienerc}, which show the detrimental effect of $H_{1}$ on entangled systems (types (i) and (ii) respectively). Similar to Fig. \ref{fig:110000_nnnienerc}, the loss in EoF scales as an exponential in $N$ with Gaussian damping in the noise parameters. Again, the influence of $H_{1}$ has been averaged over 200 random realization using random numbers $\epsilon_{i}$ from a flat distribution, such that $\epsilon_{i}=\epsilon J_{max} d_{i}$ and $0 \le d_{i} \le 1$.

Figs. \ref{fig:10+01_nnnienerc} (b) and \ref{fig:gate_nnnienerc} (b) also show that for $\epsilon \le 0.1$, EoF close to unity can still be achieved for all chain lengths considered. However, we note that for larger values of $E$, chains with type (ii) entanglement suffer significantly more entanglement loss than those of type (i) which are already entangled at $t=0$.

\begin{figure}
 \centering
 \includegraphics[width=0.45\textwidth]{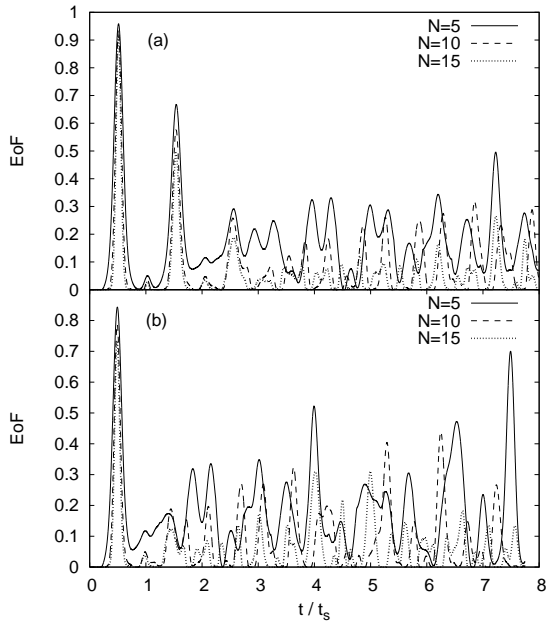}
\caption{Spin chains with input state $|\psi_{in}\rangle = 1/2(|0_{1}0_{N}\rangle+|1_{1}0_{N}\rangle+|0_{1}1_{N}\rangle+|1_{1}1_{N}\rangle) \otimes |0_{2} \cdots 0_{N-1}\rangle$:\\(a) EoF of $|\psi_{in}\rangle$ vs rescaled time $t/t_{S}$ for $\Delta=0.05$: the periodicity of the entanglement of formation is completely lost after the second peak at $t=1.5 t_{S}$.\\(b) EoF of $|\psi_{in}\rangle$ vs rescaled time $t/t_{S}$ for $\Delta=0.1$: the periodicity of the entanglement of formation is completely lost after the first peak at $t=0.5 t_{S}$.}
 \label{fig:nnnicon}
\end{figure}

As with unentangled states, next-nearest neighbor interaction is a relevant issue for entangled states. In Fig. \ref{fig:110000nnni}, we showed that a 6-spin chain suffered serious fidelity loss ($\sim 20\%$) for relatively small $\Delta = 0.01$ after about 4 periods, but still reached a fidelity of 0.9966 at $t_{M}$; similarly we see in Fig. \ref{fig:gate_nnnienerc} that for the same value of $\Delta$, a 6-spin chain with case (ii) entanglement would reach EoF of 0.99 at $t=t_{M}$ and performs thus equally well. EoF of almost unity persists for all $N$ considered and small $\Delta$. More than 90\% of the EoF is maintained even for $\Delta$ as large as 0.05 for long chains with type (ii) entanglement while chains with type (i) suffer significantly more and long chains lose over 30\% of their EoF for the same value of $\Delta$. We note that for values of $\Delta \gtrsim 0.05$, in chains with type (ii) entanglement, whilst EoF is fairly well maintained at $t=t_M$, \textit{its subsequent periodicity is completely lost for any chain length} (Fig. \ref{fig:nnnicon}, for a demonstration of the same effect in an unentangled chain see Fig. \ref{fig:NNNIcomp}).

\begin{figure}
 \centering
 \includegraphics[width=0.45\textwidth]{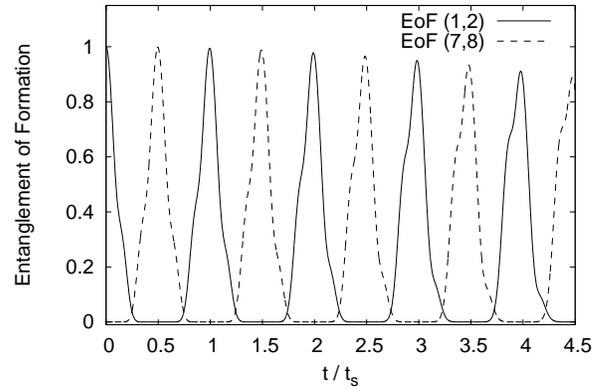}
 \caption{Influence of $H_{3}$ (Eq. (\ref{hami2})) with $\Delta=0.01$ on the entanglement of formation between the end spins of an 8 spin chain with input state $|\psi_{in}\rangle = 1/\sqrt{2}(|10000000\rangle+|01000000\rangle)$ vs. rescaled time $t/t_{S}$. The first peak at $t=0.5t_{S}=t_{M}$ is 0.9986.}
 \label{fig:0110nnni}
\end{figure}

\begin{figure}
 \centering
 \includegraphics[width=0.45\textwidth]{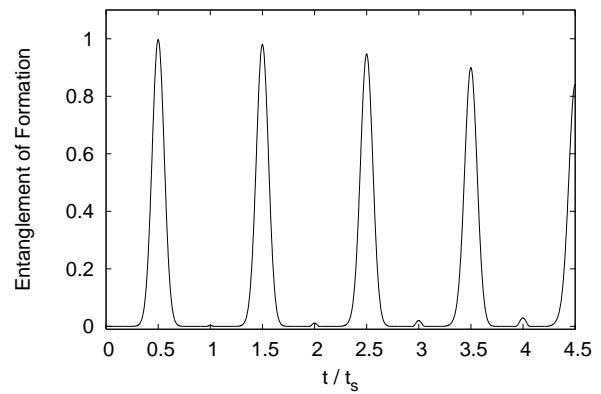}
 \caption{Influence of $H_{3}$ (Eq. (\ref{hami2})) with $\Delta=0.01$ on the entanglement of formation between the end spins of a linear 8 spin chain with input state $|\psi_{in}\rangle = 1/2(|0_{1}0_{8}\rangle+|1_{1}0_{8}\rangle+|0_{1}1_{8}\rangle+|1_{1}1_{8}\rangle)\otimes |0_{2} \cdots 0_{7}\rangle$ vs. rescaled time $t/t_{S}$. The first peak at $t=0.5t_{S}=t_{M}$ is 0.9976.}
 \label{fig:gatennni}
\end{figure}

As noted above the effect of $H_{3}$ on entangled states is different from case (i) to case (ii). In Fig. \ref{fig:0110nnni} we see that for $\Delta=0.01$ a state that is initially already entangled does not suffer very much and retains an EoF of over 90\% at $t \approx 4 t_{S}$, after 4 periods, whereas a chain where the entanglement is created through the system dynamics suffers a loss in EoF of about 15\% after 4 periods, at $t \approx 4.5 t_{S}$, as shown in Fig. \ref{fig:gatennni}.

Overall, we observe that the perturbative influence of next-nearest neighbor interaction (Eq. (\ref{hami2})) is the main cause for loss in fidelity in both unentangled states as well as states whose entanglement results only from their dynamics and which are initially unentangled, although unentangled chains suffer slightly more. Entangled chains with an initially entangled state on the other hand have been shown to be more robust against this type of defect.

Despite these variations, our study clearly demonstrates that next-nearest neighbor interactions are the most damaging form of perturbation overall, as seen in state transfer fidelity or EoF. The reason for this is that the next-nearest neighbor interaction is the only fabrication defect or limitation in the set (a)-(d) that effectively opens up new ‘channels’ for the system dynamics. The Hamiltonian $H_{3}$ of Eq. (\ref{hami2}) connects chain sites which would otherwise be disconnected (at the same order in perturbation). It therefore facilitates a more efficient (in a detrimental sense) ‘spread’ of excitations. The general consequence of this is more damage to transfer fidelity or EoF, when compared to defects (a)-(c) with the same level of noise. Numerical simulations (not shown) support this explanation, as addition of new perturbative ‘channels’ for the dynamics ‘by hand’ (as opposed to via $H_{3}$) into the full Hamiltonian can lead to similar results to those in, for example, Fig. \ref{fig:nnnicon}. It has been shown in Ref.~\cite{kay2006} that opening new channels even beyond next-nearest neighbor interaction can be compensated for if local control within the chain is possible by adjusting the nearest neighbor coupling. This degree of local control may not always be available and/or might not be desirable, so in our work we consider the longer range interactions as a potential perturbation on nearest neighbor systems designed to produce PST.

\section{Effect of non-synchronous and imperfect input operations}

\begin{figure}
 \centering
 \includegraphics[width=0.445\textwidth]{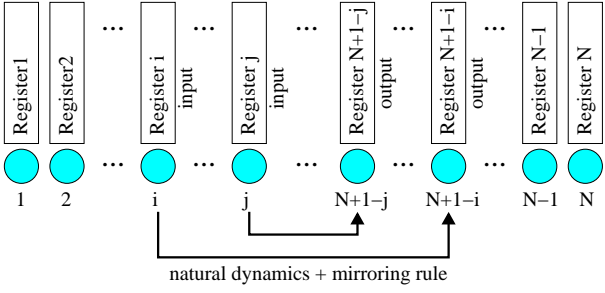}
 \caption{sketch of possible multiple qubit input/output device. In the example each spin correspond to a specific register, while the active input registers are the $i$ and $j$ registers and the information 
is transferred to the $N-i+1$ and $N-j+1$ via the mirroring rule.}
 \label{diag}
\end{figure}

In this section we will discuss the effect of imperfect excitation injection into the device when multiple excitation states are considered. A possible device configuration for input/output of multiple excitations is sketched in Fig. \ref{diag}, where each site in the chain is associated with a register which can act as input/output device. We assume that there exists a clock to which the machinery at both ends of the chain have reference. (Without such a clock even simple PST could not operate, as extraction has to be timed with respect to injection.) The timing errors we consider are with respect to this reference clock.

We will first consider the case of unentangled input states $|\phi_{\{i\},n}\rangle$ and then analyze the effects on entangled input states.

\subsection{Unentangled states: non-synchronous injection}

Let us consider the injection and transfer of a state $|\phi_{\{i\},n}\rangle$. An important question to ask is how and to what extent a time delay in input operations would alter the transmission fidelity. This question is particularly important for multiple excitations: for a single excitation a delayed injection would not alter the overall state evolution, but there would be merely an overall time shift. For multiple excitations however, it may occur that during the preparation of a state $|\phi_{\{i\},n}\rangle$, with $n>1$, not all input sites are accessed at exactly the same time. As a consequence a system which is supposed to be prepared in a two-excitation state may start evolving as a one excitation system if the two required excitations are not injected in synchrony. \\Let us focus on the latter case. When considering such a delayed input there is a finite probability that the second spin $k$ we want to inject an excitation into is already occupied. The result of this scenario is dependent on the injection mechanism, so we will consider the two main possibilities: (I) spin excitation via a Rabi-flopping control pulse on spin $k$ -- applicable for example to systems in which excitations correspond to ground state excitons confined within a quantum dot chain or to flipping the spin of an electron already confined within the chain -- and (II) injection via SWAP operation or injection of an additional particle in the chain. The latter may correspond e.g. to the scenario in which the state in the qubit of register $k$ closest to the spin chain (see Fig. \ref{diag}) is swapped with the state in the chain site $k$ -- e.g. via a train of laser pulses  in the case of exciton qubits (see \cite{spiller}) -- or to the scenario in which the main computation occurs via coherent electron transport in quantum wires (such as in \cite{bertoni}), each connected to a spin chain site.

\begin{figure}
 \includegraphics[width=0.45\textwidth]{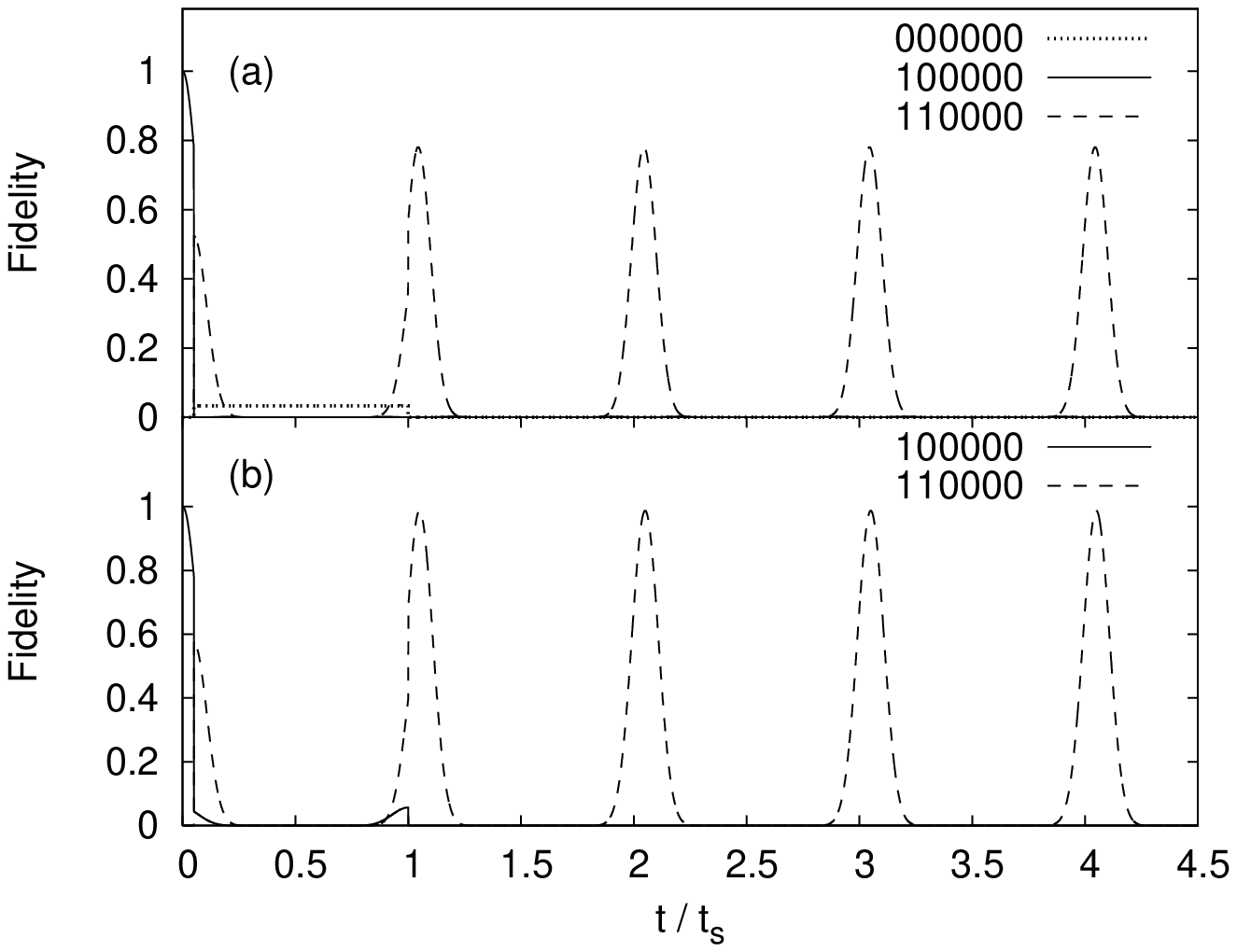}
 \caption{Fidelity with input of the second excitation in a 6-spin chain with $0.1t_{M}$ delay vs. rescaled time $t/t_{S}$:\\(a) shows injection by Rabi-flopping, where the resulting error remains in the system as the zero vector $|000000\rangle$. At $t=t_{S}$ the state of the first spin is measured and an excitation is found. This allows for refocusing the system in the two excitation subspace only. The peak of $|110000\rangle$ at multiples of $t_{S}$ plus delay is 0.7807. If at $t=t_{S}$ the second spin is measured instead, the peak fidelity becomes 0.7203.\\(b) shows injection by SWAP operation where the error remains in the system in the one excitation subspace. At $t=t_{S}$ the state of the register is measured and no excitation is found. This projects the system in the two excitation subspace only and disentangles the chain from the register. The peak of $|110000\rangle$ at multiples of $t_{S}$ plus delay is 0.9870.}
  \label{fig:delay_6_RS}
\end{figure}

In case (I) injecting an excitation corresponds to applying a $\pi$-pulse using the Hamiltonian $H_R= \Omega(t)|0\rangle \langle1|_k+\Omega^*(t)|1\rangle \langle0|_k$ with $\Omega$ the Rabi frequency (we assume that qubits can be manipulated on an individual basis). Accordingly, at the delayed injection of the second excitation, the system state will evolve as
\begin{equation}
\sum_{j}c_{\{i_j\},1}|\phi_{\{i_j\},1}\rangle\to \sum_{j\ne k}d_{\{i_j,i_k\},2}|\phi_{\{i_j,i_k\},2}\rangle+d_{0}|\phi_{\{\},0}\rangle\label{evol_1}
\end{equation}
with $d_{\{i_j,i_k\},2}=c_{\{i_j\},1}$ for $j\ne k$ and $d_{0}=c_{\{i_k\},1}$. Here we have explicitly displayed the set of indices $\{i\}$. The last term in Eq.~(\ref{evol_1}) corresponds to the error induced by having a non-zero probability $|c_{\{i_k\},1}|^{2}$ of an excitation present in spin $k$, and translates, after injection, in the probability of having {\it no-excitations} at all present in the chain (see Fig. \ref{fig:delay_6_RS}(a)). If the desired dynamics is such that a certain site $j$ has unit probability to contain an excitation at a later time $\bar{t}$ though, the system can be refocused by measuring site $j$ at $\bar{t}$: a result of ``excitation present" would collapse the system state into two-excitation dynamics, which is the closer to the desired dynamics the smaller the delay; a result of ``no-excitation present" would imply that the chain contains indeed no excitation at all. We underline that the latter result could be used as a protocol to reinitialize the chain itself.

In case (II) the scenario is very different. Should the first excitation already have a non-zero probability of occupying site $k$, the second 
excitation will have a related probability of remaining in the register. The register would then become entangled with the spin chain. The injection process can in fact be described as follows (where we assume that the presence of the excitation in site $k$ is the only cause of failed injection)
\begin{eqnarray}
& &\sum_{j}c_{\{i_j\},1}|\phi_{\{i_j\},1}\rangle\otimes |1\rangle_{register}\to \nonumber \\ 
& &\sum_{j\ne k}d_{\{i_j,i_k\},2}|\phi_{\{i_j,i_k\},2}\rangle\otimes |0\rangle_{register}+ \nonumber \\ 
& & c_{\{i_k\},1}|\phi_{\{i_k\},1}\rangle\otimes |1\rangle_{register}\label{evol_2},
\end{eqnarray}
with $d_{\{i_j,i_k\},2}=c_{\{i_j\},1}$ for $j\ne k$.

The simplest way to destroy this unwanted entanglement  is to measure, after injection, whether the second excitation is still present in the register or wire.  By this measure, we remove the entanglement between register and device, but we also get to know exactly what state the device itself is in: if the measure outcome is ``no excitation in the register" the injection has been successful and the spin chain now follows a two excitation evolution which is the closer to the desired dynamics the smaller the injection delay has been. This is described in Fig. \ref{fig:delay_6_RS}(b). If the outcome of the measurement is ``excitation in the register" the chain will continue to evolve as a one excitation system. However in the latter case we have also re-set the chain ready for trying again the injection of the second excitation at the earliest convenient time.

\begin{figure}
 \includegraphics[width=0.45\textwidth]{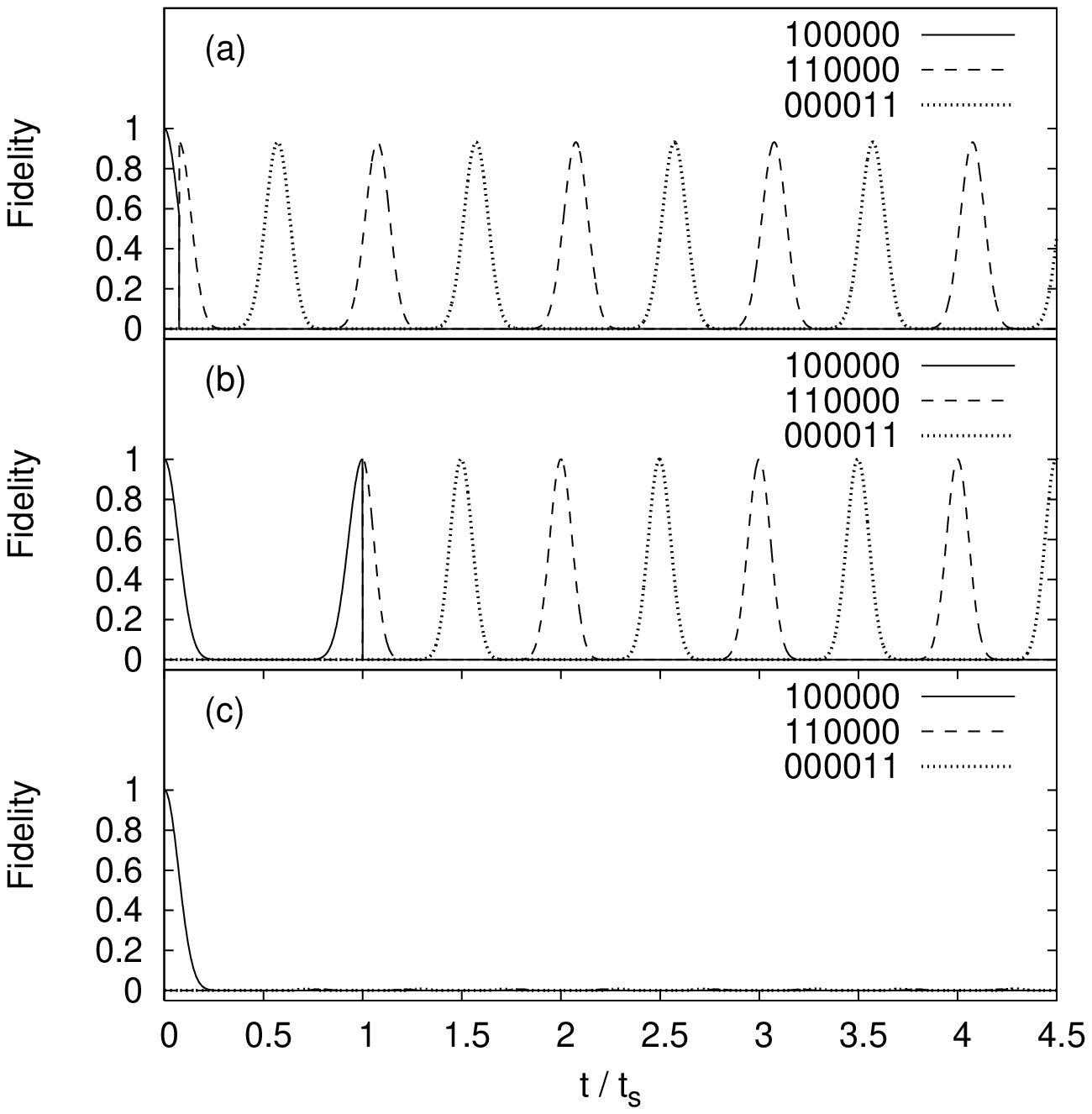}
 \caption{Effects of delayed input of a second excitation on a 6-spin chain with SWAP operation given by fidelity vs. rescaled time $t/t_{S}$.\\(a) input delayed by $0.15 t_{M}$.\\(b) input delayed by an integer number of $t_{S}$ (here: $1 t_{S}$).\\(c) input delayed by an odd multiple of $t_{M}$ (here: $1 t_{M}$).\\For (a) the maximum recurring fidelity of $|000011\rangle$ is 0.9313, at odd integer multiples of $t_M$ plus the delay. For (b), due to the complete period delay, $|000011\rangle$ emerges with unit fidelity, whereas for (c) the occurrence is negligible.}
  \label{fig:6spin2exdelays}
\end{figure}

\begin{figure}
 \includegraphics[width=0.45\textwidth]{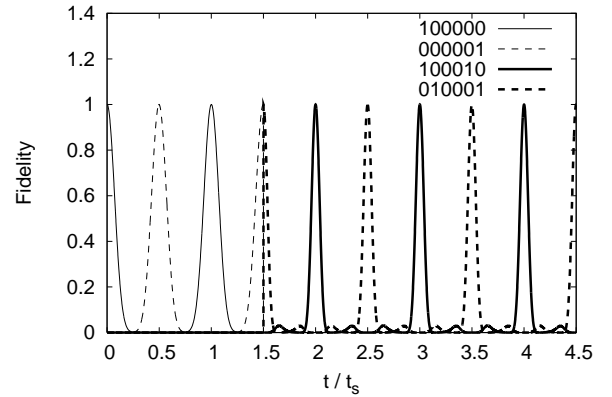}
 \caption{Fidelity after second input to a 6-spin chain is delayed by an odd multiple of $t_{M}$ (here: $3 t_{M}$) vs. rescaled time $t/t_{S}$. The peak of $|010001\rangle$ at $t=2.5t_{S}$ reaches unity.}
  \label{fig:newborn}
\end{figure}

In Fig. \ref{fig:6spin2exdelays}, we give an overview of various possible delay scenarios, assuming input by SWAP operation with error correction by measurement of the injecting register or wire performed immediately after the attempted injection of the second excitation. In panel (a), the delay of the second excitation is equal to $0.15 t_{M}$. Even with the error correction measurement we see an impact on the system as the delay between the two injections puts a cap on the fidelity of the revivals of the desired two-excitation input state. The larger the delay (up to the mirror time $t_{M}$), the more serious the impact. Frame (b) demonstrates that even if timely simultaneous injection were not possible, the ability to inject the second excitation an integer number of periods $t_{S}$ after the first one would allow for perfect revivals. In contrast, the third frame (c) shows that a second injection exactly $t_{M}$ away from the ideal case would result in the total decay of the desired state. However, we have to keep in mind that at $t=t_{M}$ where $|\phi_{\{i\},1}\rangle$ has the lowest fidelity, the corresponding twin state will be achieving perfect fidelity and that therefore a new state will emerge, we illustrate this in Fig. \ref{fig:newborn}: at a time equal to an odd multiple of $t_{M}$, the initial input state $|100000\rangle$ has zero fidelity, while its twin state $|000001\rangle$ has unit fidelity. Injection of the second excitation into the second site (which would have resulted in the ideal state $|110000\rangle$ if there had been no delay) results in $|000001\rangle \rightarrow |010001\rangle$. This state then continues to evolve with its twin state $|100010\rangle$, both of them alternately reaching perfect fidelity. The additional features on the plots of vectors with two excitations, as well as the narrowing of the main peak are due to excitation of the particular superpositions of energy eigenstates in the two-excitation subspace that correspond to our initial state. This shows how we can use delayed input to transfer states which are different from the ``twin'' sites of the input sites while still assuring PST. This may be useful if e.g. not all sites are accessible for input operations.

\subsection{Unentangled states: imperfect input operations}

Imperfect injection can occur even without delays between different excitations' injections due to other device imperfections. A typical example could be the possibility that in the SWAP injection scenario one of the two excitations is partially reflected into the wire. This may occur in a device in which the spin chain is formed by gate-defined quantum dots where electrons are injected via (computational) wires, with the reflection being caused by  imperfect lowering of a potential barrier between the wire and the selected spin chain site $k$. In this case the injection would be described by
\begin{eqnarray}
& &|\phi_{\{ \},0}\rangle\otimes |1\rangle_{register_{k}}\otimes |1\rangle_{register_{j}}\to \nonumber \\ 
& &d_{\{i_j,i_k\},2}|\phi_{\{i_j,i_k\},2}\rangle\otimes |0\rangle_{register_{k}}\otimes |0\rangle_{register_{j}}+ \nonumber \\ 
& & c_{\{i_j\},1}|\phi_{\{i_j\},1}\rangle\otimes |1\rangle_{register_{k}}\otimes |0\rangle_{register_{j}}\label{evol_3},
\end{eqnarray} where the first spin has been perfectly injected at site $j$, while there is a reflection probability $|c_{\{i_j\},1}|^2$ associated with spin $k$. In this scenario measuring the absence (presence) of excitation in the register after injection would ensure that the chain is undergoing exactly the desired dynamics (or that the chain is ready for re-injecting the second excitation at the closest suitable time). The above scenario can be straightforwardly extended to the case in which a finite probability of reflection is associated with both injection sites.\\
While in this paper we are only discussing delays at the input stage, similar problems may arise of course at read-out, should the extraction of a state covering multiple sites not be as timely as we would hope it to be. It is also worth noting that any peak following delayed input is shifted forward in time linearly with increasing delay.

\subsection{Entangled states}

\begin{figure}
 \centering
 \includegraphics[width=0.45\textwidth]{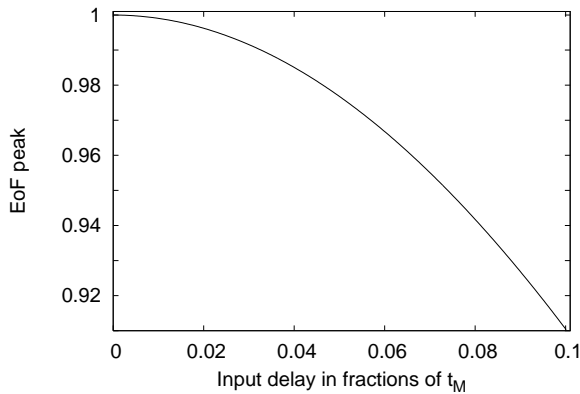}
 \caption{Maximum value of the first entanglement of formation peak at $t_{M}$ vs. input delay for an 8-spin chain with initial input state $|\psi_{in}\rangle = 1/2(|0_{1}0_{8}\rangle+|1_{1}0_{8}\rangle+|0_{1}1_{8}\rangle+|1_{1}1_{8}\rangle)\otimes |0_{2} \cdots 0_{7}\rangle$ for both Rabi flopping and SWAP operation type of injections.}
 \label{fig:gatedelay}
\end{figure}

When considering imperfect or non-synchronous input and entanglement transfer, we see a similar effect to that on unentangled states. Again, as the overall desired system state is more complicated, we monitor the evolution of the amount of entanglement in the relevant sub-system, instead of individual vectors. In Fig. \ref{fig:gatedelay} we show the evolution of the first EoF peak after the delayed input has taken place, for case (ii) where entanglement is created from an initial product state. In this setting it is not possible to implement the refocusing protocol for the Rabi flopping, but we are assuming that the confirmed injection protocol for the SWAP operation is implemented instantaneously, just after the injection attempt. Due to the fact that the two injection sites in the state described in this figure are at opposite chain ends, there is virtually no difference between injection by Rabi flopping and injection by SWAP operation. As long as the second injection is only delayed by a small fraction of $t_{M}$, the first excitation will not have spread out far enough yet to affect the second injection site. We see that even for delays as large as $0.1 t_{M}$, the system retains over 90\% of its possible entanglement.

\begin{figure}
 \centering
 \includegraphics[width=0.45\textwidth]{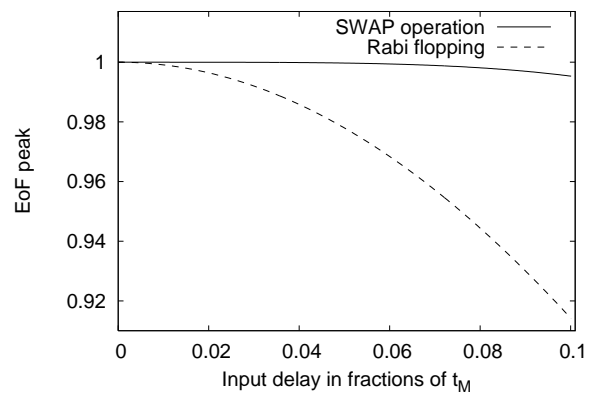}
 \caption{Maximum value of the first entanglement of formation peak vs. input delay for a 6-spin chain with initial input state $|\psi_{in}\rangle = 1/\sqrt{2}(|100000\rangle +|010000\rangle)$ for both SWAP operation and Rabi flopping type injections.}
 \label{fig:0110delay}
\end{figure}

When we consider two adjacent injection sites, there is instead a clear discrepancy between the decay in entanglement of formation in a chain with input via Rabi flopping or SWAP operation (Fig \ref{fig:0110delay}). Again, we assume that there is an instantaneous SWAP correction protocol applied, while refocussing after the Rabi-flopping is not possible. As in the case of non-synchronous input in unentangled states, a delay in a SWAP type injection perturbs the system far less than using Rabi flopping, and shows virtually no loss in the amount of entanglement in the system, even for delays as large as $0.05t_{M}$. If instead Rabi flopping is used, the decay in EoF is similar to that of Fig. \ref{fig:gatedelay}.

\section{Conclusions}

In this paper we have considered a variety of physically relevant perturbative factors in spin chains. We have investigated in detail their effects on information transfer and entanglement generation and transfer. For different forms of perturbation, the results of our extensive numerical studies on the quality of transfer can be captured with a straightforward analytic expression that demonstrates exponential damping with the number of spins on the chain and Gaussian dependence on the relevant perturbation parameter. This expression provides a simple tool for estimating the efficiency of a chain under the action of perturbations. We have also related our dimensionless perturbation parameter scales to actual parameters for candidate spin chain realizations such as quantum dots with exciton qubits and graphene dots with spin qubits, providing a calibration of our estimator tool for these experimental systems.

We have considered the transport of both unentangled states as well as entangled states via spin chains, subject to various forms of perturbation, and a general conclusion is that next-nearest neighbor interactions are the most damaging to transfer fidelity or EoF. The reason for this
can be traced back to the fact that the next-nearest neighbor interaction is the only perturbation that effectively opens up new `channels' for the system dynamics. By connecting sites which would otherwise be disconnected (at the same order in perturbation), it allows a more efficient (in a detrimental sense) `spread' of excitations, and consequently drastically diminishes the occurrence of quantum coherent effects such as revivals and PST. In particular, we have seen that the introduction of next-nearest neighbor coupling may lead to non-negligible quality loss after only a few periods. Systems with simpler input states are generally slightly less affected than those of a slightly more complex nature but as different perturbation factors affect different excitation subspaces, there is no clear advantage of one particular type of state in terms of robustness. However, for all fabrication defects considered we have found that the transport during the first period remains of high quality for perturbation amplitudes of the order of few percent, while the periodicity itself may be destroyed for next-nearest neighbor couplings above about $5\%$. This is of particular concern for schemes based on dipole-dipole interactions, where, for roughly equidistant sites, the next-nearest neighbor interactions are of the order of $10\%$.
Nevertheless, due to the Gaussian dependence of transport efficiency on perturbation amplitudes, provided that the perturbations are kept below these few percent thresholds, spin chains are demonstrated to be very good candidates for the implementation of solid state quantum information processing devices, robust against various forms of perturbation.

Additionally, we have seen that the effect of imperfect input or injection operations leads to a permanent (but constant) loss in information transport quality and entanglement generation. For input delayed by a time approaching the mirroring time $t_{M}$ the transport of the intended state is replaced by a new input state, which is then subject to the system dynamics as per usual. Furthermore, there is a fundamental difference between the two input methods we have considered. If Rabi flopping is used and the injection of the second excitation is delayed, the error induced by the imperfect input remains in the system, except if a site is expected to be in state $|1\rangle$ at a known time, when measurement of this site can recover dynamics close to the ideal case. In the case of input by SWAP operation, the error can be dramatically reduced by subsequent measuring of the environment. When considering SWAP type injection, unentangled states and type (i) entanglement states are remarkably robust against non-synchronous injection, with hardly any loss in fidelity even for large delays. Type (ii) dynamically-generated entangled states are instead more affected due to their more complex set-up. Even so the loss in EoF remains less than 10\% for delay values up to 10\% of the mirroring time $t_{M}$.

Our studies demonstrate quantitatively the criteria that need to be met, in terms of perturbation scales and injection errors, for imperfect spin chains to work efficiently as quantum information transfer and entanglement transfer/generation devices. With modest errors at or below the few percent level, spin chains prove to be good and robust devices, which is very encouraging for future experiments on these systems.
\\
RR was supported by EPSRC-GB and HP. IDA acknowledges partial support by HP. IDA and RR acknowledge the kind hospitality of the HP Research Labs Bristol.


\end{document}